%% file: brahim10.tex
\documentstyle[aps,prl]{revtex}

\begin{document}


\title{Non--linear supersymmetric $\sigma$--Model for Diffusive
  Scattering of Classical Waves with Resonance Enhancement}
\author{B. Elattari$^{1,2}$, V. Kagalovsky$^{1}$, and 
H.A. Weidenm\"uller$^{1}$}
\address{$^1$Max-Planck-Institut f\"ur Kernphysik, 69029 Heidelberg, 
Germany;\\
$^2$Universit\'e Choua\"\i b 
Doukkali,  Facult\'e des Sciences, El Jadida, 
Morocco}

\date{\today}
\maketitle

\begin{abstract}We derive a non--linear $\sigma$--model for the
  transport of light (classical waves) through a disordered medium. We
  compare this extension of the model with the well--established
  non--linear $\sigma$--model for the transport of electrons
  (Schr\"odinger waves) and display similarities of and differences
  between both cases. Motivated by experimental work (M. van Albada et
  al., Phys. Rev. Lett. {\bf 66} (1991) 3132), we then generalize the
  non--linear $\sigma$--model further to include resonance scattering. 
  We find that the form of the effective action is unchanged but that 
  a parameter of the effective action, the mean level density, is
  modified in a manner which correctly accounts for the data. 
\end{abstract}


The supersymmetric non--linear $\sigma$--model (SNSM) \cite{Efe83} is
a powerful tool for the description of electron transport through
disordered media. In compact form, SNSM contains information on diffusion,
reflection and transmission, including non--perturbative aspects like
localization and parametric correlations. The purpose of this letter
is twofold. (i) We extend SNSM to the transport of light (classical
waves) through disordered media. The extension shows that SNSM is not
restricted to Schr\"odinger waves but applies equally to classical
waves of sufficiently high frequency. We use the Ward identities to
check our results. These identities differ in form significantly for
classical and for Schr\"odinger waves. Although the effective action
of the SNSM is the same in both cases. the Ward identities are
fulfilled because the source terms differ. (ii) We generalize SNSM
further by allowing for both diffusive and resonance scattering. We do
so with the help of random matrix theory. The generalization is
motivated by an experiment on resonant light scattering \cite{Alb91}. 
The generalization leaves the {\it form} of the effective action
unchanged but endows a parameter of the action (the mean level
density) with an energy dependence which correctly accounts for the
data. As a result, we obtain a unified description of diffusive and of
resonant scattering for both Schr\"odinger waves and classical waves.

We first derive SNSM for classical waves and then turn to the
generalization involving resonance scattering. Following common
practice \cite{reviewlight}, we replace vector fields by scalar
quantities and consider the scalar wave equation 
\begin{equation}
[\Delta +k^2\epsilon ({\bf r})]\Phi =0
\label{wav}
\end{equation}
where $k$ is the wave number. The dielectric constant $\epsilon ({\bf
  r})$ is the sum of the background dielectric constant $\epsilon_0$
(a constant) and the fluctuating part $\delta\epsilon ({\bf r})$ which
represents disorder. We assume that $\delta\epsilon ({\bf r})$ is a
Gaussian random process with vanishing first moment and a second
moment given by
\begin{equation}
<\delta\epsilon ({\bf r_1})\delta\epsilon ({\bf r_2})>=
\frac{4\pi}{\ell k^{d+1}}\delta ({\bf r_1}-{\bf r_2}),
\label{diel}
\end{equation}
where $\ell$ is the elastic mean free path (connected to the diffusion
constant $D$ and the energy transport velocity $v_E$ via $D = (1/d)v_E
\ell)$, and $d$ is the dimension of the system. It is well to recall
the difference between the scalar wave equation~(\ref{wav}) and the
Schr\"odinger equation for an electron with energy $E$ in a disorder
potential $V$. Formally, the quantity $k^2 \epsilon_0$ in
Eq.~(\ref{wav}) corresponds to $E$, while $k^2 \delta \epsilon$ plays
the role of $V$. We note, however, that $k^2 \delta \epsilon$ is
energy dependent while $V$ is not. This is the fundamental difference
between classical and Schr\"odinger waves. 

Following Efetov \cite{Efe83}, we calculate the supersymmetric
generating functional for the average of the product of a retarded and
an advanced Green function for the differential operator in
Eq.~(\ref{wav}). We mention in passing that John and Stephen
\cite{john} derived a non--linear sigma model for classical waves. 
This derivation was confined, however, to waves of fixed energy and
thus bypassed the crucial issue of correlations between amplitudes at
{\it different} energies. Let $k_1, k_2$ be the $k$--values of the two
Green functions. With $k_0^2 = (k_1^2 + k_2^2)/2$ and $\Delta k^2 = k_1^2 -
k_2^2$, the effective action is given by  
\begin{equation}
{\cal L}[Q]=\int d{\bf r} \left(
\frac{\pi\nu}{8\tau}{\rm trg}Q^2+\frac{1}{2}{\rm trg}
\log\left[ k_{0}^{2}+\frac{\Delta k^2}{2}{\rm L}+\Delta +i\eta {\rm L}-
\frac{1}{2\tau}Q\left( 1+\frac{\Delta k^2}{2k_{0}^{2}}{\rm
  L}\right)\right] \right).
\label{action}
\end{equation}
In order to facilitate a comparison with Efetov's expression for
electrons. we have introduced the density of states $\nu$ per unit of
$k_{0}^{2}$ and per unit of volume, and the quantity $\tau =
k_{0}^{d-3}\ell/(2\pi ^2\nu )$ which is formally equivalent to the
mean free time. We use the notation of ref. 
\cite{Ver85}. The symbol trg denotes the supertrace. The supermatrices
have dimension eight. The matrix ${\rm L}$ is given by
diag$(1,1,-1,-1,1,1,-1,-1)$. The term $\Delta {\cal L} = (1 / 2 \tau)
Q {\rm L} ( \Delta k^2 / 2 k_0^2)$ is due to the energy dependence of
the scattering potential $k^2 \delta \epsilon$ and would be absent in
the case of electrons. Except for this term the effective action
formally agrees with the one derived for electrons. The non--linear
$\sigma$--model is generically derived with the help of a
saddle--point condition that is valid for weak disorder, $k_0 \ell \gg
1$. Comparing the two terms proportional to $\Delta k^2$ in
Eq.~(\ref{action}), we see that for weak disorder, the 
term $\Delta {\cal L}$ can be neglected. Then, ${\cal L}[Q]$ is
identical to the case of electrons. The saddle--point condition has
the solution $Q=i{\rm L}$. Expanding around the saddle point and
integrating over the massive modes, we obtain exactly the same form
for the effective action as in the case of electrons. Hence, {\it
  there is no difference between the non--linear sigma models for
  Schr\"odinger waves and for classical waves}. This is the first
important result of our work. It obviously extends to the generating
functionals of all higher correlation functions and, thus, applies
universally. The actual differences between the two theories are due
to the different forms of the source terms.

The Ward identities for classical and for Schr\"odinger waves differ
substantially~\cite{reviewlight}. Because of the frequency dependence
of the $k^2 \delta \epsilon$ term in Eq.~(\ref{wav}), there occurs an
additional term for classical waves which is absent for Schr\"odinger
waves. How can identical effective actions give rise to different Ward
identities? This is due to the difference of the source terms of both
theories. The Ward identities are indeed fulfilled in both
cases~\cite{rour1}. 

We turn to the generalization of SNSM including resonance scattering. 
To motivate both the problem and our modeling, we recall the
experiment by van Albada et al.~\cite{Alb91}. For light scattering,
disorder is often produced artificially: A powder is immersed into
some liquid. Random scattering occurs if the dielectric constants of
powder and liquid are sufficiently different. In the experiment of
ref.~\cite{Alb91}, the powder used (TiO$_2$) consisted of grains with
a size distribution centered around a diameter of $220$ nm. For such
grains, a Mie resonance occurs close to the wavelength $\lambda \sim
630$ nm of the laser light used in the experiment, causing a resonance
enhancement of the scattering. This enhancement led to unusually low
values of the diffusion constant $D$. Indeed, with $D$ determined from
the intensity autocorrelation function versus frequency of the
transmitted light. and the elastic mean free path $l$ determined from
weak localization (enhanced backscattering), or from the dependence of
the transmitted intensity on the length $L$ of the disordered slab,
the relation $D = 1/3 \ v_E l$ yielded a value $v_E = (5 \pm 1) \
10^7$  m s$^{-1}$ for the energy transport velocity $v_E$ through the
disordered medium. This value is about an order of magnitude smaller
than the phase velocity.

This surprising result has been understood both for small concentration
\cite{Alb91,KoganK,Cwi,Kroha,Barab,vanT93,vanTE93} and, more recently,
also for strong concentration \cite{Bus95,Bus96} of the scatterers. (In
the latter case, the resonant structure in the diffusion constant versus 
frequency disappears, and an overall decrease of $D$ is observed
\cite{Gar}). Qualitatively speaking, the transport velocity is reduced
because on its way through the medium, the energy is stored for some
time in the Mie resonances. 

Our generalization of the SNSM goes beyond this work in two ways. 
First, it yields a unified theoretical framework in which both average
and fluctuation properties can be calculated on the same footing. This
is in contrast to previous approaches \cite{Alb91,Bus95,Bus96} which
use the Bethe--Salpeter equation \cite{Alb91} or a mean--field
approximation \cite{Bus95,Bus96} for the calculation of the transport
velocity $v_E$, and a diagrammatic impurity perturbation expansion for
the intensity autocorrelation function. Second, the generalized SNSM
identifies the energy dependence of the mean level density $\rho(E)$
as the culprit for the observed deviation from standard behavior. A
simple argument which yields the same result as the analytical
derivation and illuminates the physical content may be helpful at this
point. In the case of electrons \cite{review}, the Thouless formula $g
= E_c / \Delta$ connects the average conductance $g$ with the Thouless
energy $E_c = \hbar D / L^2$ and the mean level density $\rho(E) = 1 /
\Delta$. Here, $L$ is the length of the sample. The presence of 
numerous resonances with equal resonance energies $E_1$ leads to a 
local Breit--Wigner--shaped increase of $\rho(E)$ near $E_1$. Since $g$
is not affected by the presence of the resonances, the Thouless
formula implies that $E_c$ and, hence, the diffusion constant have a
Breit--Wigner--like dip near $E_1$. Because of the equality of the
effective action in the SNSM for Schr\"odinger and for classical
waves, this argument applies likewise to scattering of light. With
increasing concentration of scatterers, the dip widens and becomes
less deep. Eventually, this results in an overall decrease of $D$ over
a wide frequency interval. We note that in the context of SNSM, the
relation $D = 1/3 v_E l$ is not used explicitly. It is replaced by the
Thouless relation.

To account for the presence of resonances, we proceed as follows. For
the quasi one--dimensional geometry appropriate in the present
context, we use the well--known identity between SNSM and a band
random matrix model~\cite{Iid90}. In the framework of the latter, it
is easy to model the presence of additional resonances. With standard
supersymmetry techniques, we finally map the resulting generalized
band random matrix model back onto SNSM. This yields the required
generalization. The procedure is the same for classical and for
Schr\"odinger waves.

In the {\it absence} of resonances, a quasi one--dimensional
disordered system of length $L$ for diffusive scattering of electrons
or light is modeled \cite{Iid90} as consisting of many longitudinal
slices of length $l_0 \ll L$. Within each slice, the Hamiltonian is
modeled as a matrix $H_{\rm GOE}$ of dimension $N$ belonging to the
GOE, the random--matrix ensemble with orthogonal symmetry. Neighboring
slices are coupled by Gaussian--distributed uncorrelated random matrix
elements. The strength of this coupling defines the diffusion constant
$D_0$. The first and the last slice are coupled to the channels, i.e.,
the asymptotic states for free propagation of electrons or light. The
transmission through the disordered region is given in terms of the
squares of elements of the scattering matrix which connect incident
and outgoing channels on either side of the disordered region.

To account for the presence of Mie scatterers, we modify the form of
the Hamiltonian $\bf H$ within each slice. Now, $\bf H$ is a matrix of
dimension $N + m$ where $m$ is the number of Mie scatterers within
that slice. The value of $m$ is given by the concentration of
scatterers, and by the linear dimensions of the slice. In an $(N,m)$
block representation, $\bf H$ has the form
\begin{equation}
{\bf H}=\left( \begin{array}{cc} \begin{array}[b]{c}
H_{\rm GOE} \end{array} & V \\
V^T & E_1\times I_m+H_{\rm res}
\end{array}
\right). 
\label{ham}
\end{equation}
Here $T$ denotes the transpose. The matrix $H_{\rm GOE}$ of dimension
$N$ was introduced in the previous paragraph, $I_m$ is the
$m$--dimensional unit matrix, $E_1 = \hbar \omega_1$ is the resonance
energy of each of the $m$ Mie scatterers with equal resonance
frequencies $\omega_1$, the rectangular matrix $V$ couples the $m$
resonances to $H_{\rm GOE}$, and the random matrix $H_{\rm res}$
describes the coupling between the $m$ resonances. The matrices $V$,
$H_{\rm res}$, and $H_{\rm GOE}$ are uncorrelated. The second moment
$(H_{\rm GOE})_{\mu \nu} (H_{\rm GOE})_{\mu' \nu'}$ of $H_{\rm GOE}$
is given by $(\lambda^2/N) (\delta_{\mu, \mu'} \delta_{\nu, \nu'} +
\delta_{\mu, \nu'} \delta_{\nu, \mu'})$. Here, $\lambda$ determines
the average level spacing $\Delta_S$ in each slice. For $m = 0$, we
obtain Wigner's semicircle law for the mean level density, typical for
random matrix theory. We assume the energy $E_1$ to lie at the center
of the semicircle, $E_1 = 0$ where $\Delta_S = \pi \lambda / N$. The
matrix $H_{\rm res}$ is also a member of a GOE with a second moment of
the same form as for $H_{\rm GOE}$ but with $\lambda$ replaced by
$\lambda_1$. The strength factor $\lambda_1$ is chosen in such a way
that the interaction between resonances results in a lifting of the
degeneracy which is of the order of the mean level spacing $\Delta_S(0)$ 
so that $\lambda_1 \ll \lambda$. Too strong an interaction would wash out
the resonance structure altogether. Without loss of generality, the
rectangular matrix $V$ can be taken to have non--zero elements $v$
only on the main diagonal $\nu = m$. Since all scatterers are assumed
to be identical, all these elements are taken to be equal. It is
possible to estimate $\lambda_1$ and $v$ microscopically from the
properties of the Mie resonance, and from the concentration $\mu$ of
scatterers. This is not done here. Suffice it to say that $\lambda_1$
grows more strongly than linearly with $\mu$ \cite{Bra98}. At the end
of the calculation, we take the limit $N \rightarrow \infty$ \cite{Ver85}.

Before we cite complete results, it instructive to consider the case
of a single slice. After averaging and the Hubbard--Stratonovich
transformations relating to the two different statistical ensembles
$H_{\rm GOE}$ and $H_{\rm res}$, two supermatrices are introduced,
leading to two coupled saddle--point equations. These are solved
easily. The resulting mean level density has the form
\begin{equation}
\rho (E) = \frac{N}{\lambda\pi} + m\frac{(\Gamma + b
  \lambda_1)/\pi} {(E - (E/2\lambda )\Gamma -a
  \lambda_1)^2+(\Gamma + b \lambda_1)^2}. 
\label{dos1}
\end{equation}
Here $\Gamma =v^2/\lambda$ is the width of the Lorentzian in the
absence of interresonance coupling, and $a+ib=\sigma_1$ is the
solution of the second saddle-point equation $\sigma_1 = \lambda_1 /
(E-\lambda_1\sigma_1-\Gamma E/2\lambda -i\Gamma)$. The last term shows
the resonance enhancement of $\rho$, centered at the energy $E_1 = 0$
of the $m$ resonances. The total width $(\Gamma + b\lambda_1)$ is
determined by both, the strength $\lambda_1$ of the interresonance
coupling and the strength $v$ of the coupling to the random
scatterers. For small $\lambda_1$ (corresponding to a low value of the
concentration $\mu$), we regain the expression for uncoupled
($\lambda_1 = 0$) resonances given in ref. \cite{our}, whereas in the
limit of high concentrations ($\lambda_1\gg\Gamma$) the Lorentzian
peak in $\rho(E)$ disappears (see Fig. 1). It is easy to check that
$\rho(E)$ obeys the requirement $\int \rho(E) dE = N+m$. The
calculation of the two--point function suffers from the presence of
two small parameters $E-E_1$ and $\lambda_1$ in addition to
$\Delta k^2$. However, under the realistic condition $\Gamma \gg
\Delta_{\rm eff} = (\rho)^{-1}$, the remaining supersymmetric integral
attains a simple form \cite{our}: The effective action coincides with
the case without resonance scattering (pure GOE) {\it except for a
rescaling of the average level spacing}, i.e. the replacement $\pi
\lambda / N \to \Delta_{\rm eff}$. All resonance effects follow from this
rescaling. Additional differences to the pure GOE case can occur only
in the wings of the resonance in Eq.~(\ref{dos1}). or in cases where
$\Gamma \sim \Delta_{\rm eff}$.

Using standard supersymmetry, we show that this result carries over to
the case of a quasi one--dimensional sample consisting of many slices. 
The effective action of the resulting SNSM has Efetov's form,
\begin{equation}
{\cal L}_{\rm eff} = -\frac{\pi}{8V\Delta_{\rm eff}}\int {\rm trg}
\left[ D_{\rm eff}(\nabla Q)^2+2i\Delta k^2{\rm L}Q\right]d{\bf r} \ ,
\label{eff}
\end{equation}
where the effective diffusion constant $D = D_0 \Delta_{\rm
  eff}/\Delta$ is 
given in terms of the diffusion constant $D_0$ and the mean level
spacing $\Delta$ in the absence of resonances, and in terms of the
effective mean level spacing $\Delta_{\rm eff}$ due to the presence of
resonant scatterers, and $V$ is the volume of the system. 
Eq.~(\ref{eff}) implies that the transmission
through a sufficiently long sample is unaffected by the presence of
resonances while all correlation functions as well as $D$ do
depend on the effective mean level spacing $\Delta_{\rm eff}$ and, thus,
show a strong dependence on the resonances.   

To compare with experiment, we note: (i) For classical waves in a
uniform medium in $d$ dimensions, the non--resonant part of the
density of states is proportional to $\epsilon^{d/2}$. Increasing the
number of resonating spheres, we increase the effective dielectric
constant of the medium. Hence, $\lambda$ should decrease with $\mu$. 
(ii) As pointed out above, $\lambda_1$ is expected to increase
monotonically with $\mu$. Both these trends affect the dependence on
concentration of $\Delta_{\rm eff}$ and, therefore, of $D \sim
\Delta_{\rm eff}$. 
With increasing $\mu$, point (i) causes a decrease of $D$ far away
from the resonance, and point (ii) a widening of the resonance. Fig. 1
shows $D$ versus energy for different concentrations. In agreement
with the experimental results \cite{Alb91,Gar}, we find a deep dip in
the diffusion constant at the resonance energy. This dip is caused by
the resonance enhancement of $\Delta_{\rm eff}$ and is particularly
pronounced at low concentrations of the resonant scatterers. As the
concentration increases, the dip is smeared out. At the same time, the
value of $D$ outside the resonance decreases. We stress that the
presence of $H_{res}$ in the Hamiltonian is crucial for the agreement
with experiment. 

In summary, we have shown that for sufficiently high frequency, the
effective action in the supersymmetric non--linear $\sigma$--model for
diffusive scattering of classical waves is identical in form with the
analogous expression for electron transport in disordered media. This
statement holds in spite of the fact that the Ward identities differ. 
Moreover, we have generalized the supersymmetric non--linear
$\sigma$--model to include resonance scattering. This generalization
leaves the {\it form} of Efetov's effective action unchanged but
endowes the {\it parameters} with an energy dependence of
Breit--Wigner form. As a result, the transmission remains unchanged
but correlation functions and diffusion constant show a strong
resonance dependence. This is in agreement with experiment. The
results are also valid for electrons and therefore could apply to
future experiments on mesoscopic samples.

Acknowledgments: Discussions with Dr. A. M\"uller-Groeling in the
early stages of this work were essential for the formulation of the
problem and are gratefully acknowledged. V. K. gratefully acknowledges
the support of a MINERVA Fellowship and useful discussions with
Dr. Y. Fyodorov. B. E. wishes to thank Prof. A. Nourreddine for
valuable discussions, and acknowledges the support of a
Max-Planck-Fellowship.

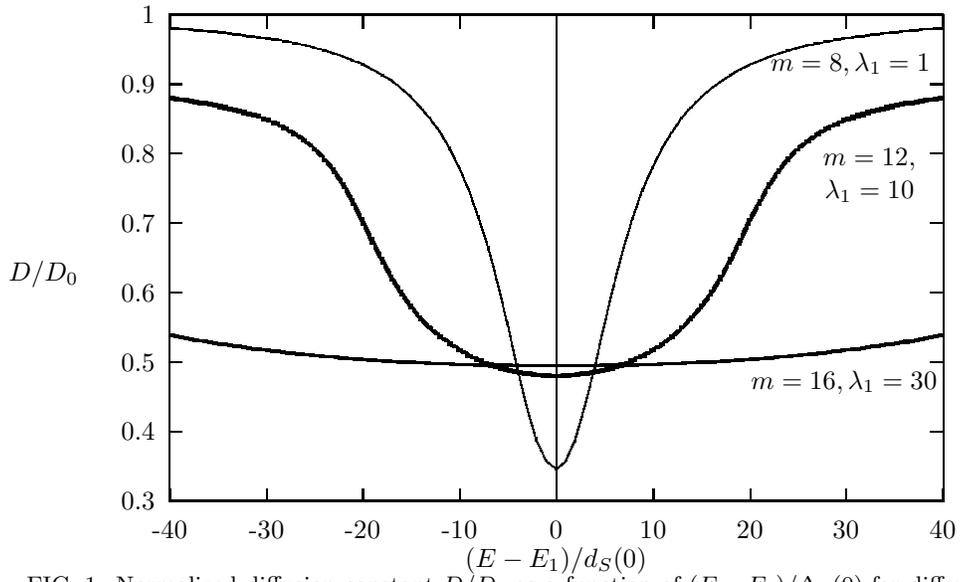
\begin{figure}
\input{Eplfig}
\caption{Normalized diffusion constant $D/D_0$ as a function 
of $(E-E_1)/\Delta_S(0)$ for different concentrations of the
resonances ($D_0$ is the diffusion constant of the medium in the
absence of resonances).} 
\end{figure}

\end{document}

%% file: Eplfig.tex
%
\setlength{\unitlength}{0.240900pt}
\ifx\plotpoint\undefined\newsavebox{\plotpoint}\fi
\sbox{\plotpoint}{\rule[-0.200pt]{0.400pt}{0.400pt}}%
\begin{picture}(1500,900)(0,0)
\font\gnuplot=cmr10 at 10pt
\gnuplot
\sbox{\plotpoint}{\rule[-0.200pt]{0.400pt}{0.400pt}}%
\put(828.0,113.0){\rule[-0.200pt]{0.400pt}{184.048pt}}
\put(220.0,113.0){\rule[-0.200pt]{4.818pt}{0.400pt}}
\put(198,113){\makebox(0,0)[r]{0.3}}
\put(1416.0,113.0){\rule[-0.200pt]{4.818pt}{0.400pt}}
\put(220.0,222.0){\rule[-0.200pt]{4.818pt}{0.400pt}}
\put(198,222){\makebox(0,0)[r]{0.4}}
\put(1416.0,222.0){\rule[-0.200pt]{4.818pt}{0.400pt}}
\put(220.0,331.0){\rule[-0.200pt]{4.818pt}{0.400pt}}
\put(198,331){\makebox(0,0)[r]{0.5}}
\put(1416.0,331.0){\rule[-0.200pt]{4.818pt}{0.400pt}}
\put(220.0,440.0){\rule[-0.200pt]{4.818pt}{0.400pt}}
\put(198,440){\makebox(0,0)[r]{0.6}}
\put(1416.0,440.0){\rule[-0.200pt]{4.818pt}{0.400pt}}
\put(220.0,550.0){\rule[-0.200pt]{4.818pt}{0.400pt}}
\put(198,550){\makebox(0,0)[r]{0.7}}
\put(1416.0,550.0){\rule[-0.200pt]{4.818pt}{0.400pt}}
\put(220.0,659.0){\rule[-0.200pt]{4.818pt}{0.400pt}}
\put(198,659){\makebox(0,0)[r]{0.8}}
\put(1416.0,659.0){\rule[-0.200pt]{4.818pt}{0.400pt}}
\put(220.0,768.0){\rule[-0.200pt]{4.818pt}{0.400pt}}
\put(198,768){\makebox(0,0)[r]{0.9}}
\put(1416.0,768.0){\rule[-0.200pt]{4.818pt}{0.400pt}}
\put(220.0,877.0){\rule[-0.200pt]{4.818pt}{0.400pt}}
\put(198,877){\makebox(0,0)[r]{1}}
\put(1416.0,877.0){\rule[-0.200pt]{4.818pt}{0.400pt}}
\put(221.0,113.0){\rule[-0.200pt]{0.400pt}{4.818pt}}
\put(221,68){\makebox(0,0){-40}}
\put(221.0,857.0){\rule[-0.200pt]{0.400pt}{4.818pt}}
\put(373.0,113.0){\rule[-0.200pt]{0.400pt}{4.818pt}}
\put(373,68){\makebox(0,0){-30}}
\put(373.0,857.0){\rule[-0.200pt]{0.400pt}{4.818pt}}
\put(524.0,113.0){\rule[-0.200pt]{0.400pt}{4.818pt}}
\put(524,68){\makebox(0,0){-20}}
\put(524.0,857.0){\rule[-0.200pt]{0.400pt}{4.818pt}}
\put(676.0,113.0){\rule[-0.200pt]{0.400pt}{4.818pt}}
\put(676,68){\makebox(0,0){-10}}
\put(676.0,857.0){\rule[-0.200pt]{0.400pt}{4.818pt}}
\put(828.0,113.0){\rule[-0.200pt]{0.400pt}{4.818pt}}
\put(828,68){\makebox(0,0){0}}
\put(828.0,857.0){\rule[-0.200pt]{0.400pt}{4.818pt}}
\put(980.0,113.0){\rule[-0.200pt]{0.400pt}{4.818pt}}
\put(980,68){\makebox(0,0){10}}
\put(980.0,857.0){\rule[-0.200pt]{0.400pt}{4.818pt}}
\put(1132.0,113.0){\rule[-0.200pt]{0.400pt}{4.818pt}}
\put(1132,68){\makebox(0,0){20}}
\put(1132.0,857.0){\rule[-0.200pt]{0.400pt}{4.818pt}}
\put(1283.0,113.0){\rule[-0.200pt]{0.400pt}{4.818pt}}
\put(1283,68){\makebox(0,0){30}}
\put(1290,800){\makebox(0,0){$m=8, \lambda_1=1$}}
\put(1280,300){\makebox(0,0){$m=16, \lambda_1=30$}}
\put(1320,650){\makebox(0,0){$m=12$,}}
\put(1320,600){\makebox(0,0){$\lambda_1=10$}}
\put(1283.0,857.0){\rule[-0.200pt]{0.400pt}{4.818pt}}
\put(1435.0,113.0){\rule[-0.200pt]{0.400pt}{4.818pt}}
\put(1435,68){\makebox(0,0){40}}
\put(1435.0,857.0){\rule[-0.200pt]{0.400pt}{4.818pt}}
\put(220.0,113.0){\rule[-0.200pt]{292.934pt}{0.400pt}}
\put(1436.0,113.0){\rule[-0.200pt]{0.400pt}{184.048pt}}
\put(220.0,877.0){\rule[-0.200pt]{292.934pt}{0.400pt}}
\put(23,470){\makebox(0,0){$D/D_0$}}
\put(828,20){\makebox(0,0){$(E-E_1)/d_S(0)$}}
\put(220.0,113.0){\rule[-0.200pt]{0.400pt}{184.048pt}}
\put(221,856){\usebox{\plotpoint}}
\put(221,854.67){\rule{3.614pt}{0.400pt}}
\multiput(221.00,855.17)(7.500,-1.000){2}{\rule{1.807pt}{0.400pt}}
\put(236,853.67){\rule{3.614pt}{0.400pt}}
\multiput(236.00,854.17)(7.500,-1.000){2}{\rule{1.807pt}{0.400pt}}
\put(251,852.17){\rule{3.100pt}{0.400pt}}
\multiput(251.00,853.17)(8.566,-2.000){2}{\rule{1.550pt}{0.400pt}}
\put(266,850.67){\rule{3.614pt}{0.400pt}}
\multiput(266.00,851.17)(7.500,-1.000){2}{\rule{1.807pt}{0.400pt}}
\put(281,849.67){\rule{3.854pt}{0.400pt}}
\multiput(281.00,850.17)(8.000,-1.000){2}{\rule{1.927pt}{0.400pt}}
\put(297,848.17){\rule{3.100pt}{0.400pt}}
\multiput(297.00,849.17)(8.566,-2.000){2}{\rule{1.550pt}{0.400pt}}
\put(312,846.17){\rule{3.100pt}{0.400pt}}
\multiput(312.00,847.17)(8.566,-2.000){2}{\rule{1.550pt}{0.400pt}}
\put(327,844.17){\rule{3.100pt}{0.400pt}}
\multiput(327.00,845.17)(8.566,-2.000){2}{\rule{1.550pt}{0.400pt}}
\put(342,842.17){\rule{3.100pt}{0.400pt}}
\multiput(342.00,843.17)(8.566,-2.000){2}{\rule{1.550pt}{0.400pt}}
\put(357,840.17){\rule{3.100pt}{0.400pt}}
\multiput(357.00,841.17)(8.566,-2.000){2}{\rule{1.550pt}{0.400pt}}
\put(372,838.17){\rule{3.300pt}{0.400pt}}
\multiput(372.00,839.17)(9.151,-2.000){2}{\rule{1.650pt}{0.400pt}}
\multiput(388.00,836.95)(3.141,-0.447){3}{\rule{2.100pt}{0.108pt}}
\multiput(388.00,837.17)(10.641,-3.000){2}{\rule{1.050pt}{0.400pt}}
\multiput(403.00,833.95)(3.141,-0.447){3}{\rule{2.100pt}{0.108pt}}
\multiput(403.00,834.17)(10.641,-3.000){2}{\rule{1.050pt}{0.400pt}}
\multiput(418.00,830.95)(3.141,-0.447){3}{\rule{2.100pt}{0.108pt}}
\multiput(418.00,831.17)(10.641,-3.000){2}{\rule{1.050pt}{0.400pt}}
\multiput(433.00,827.94)(2.090,-0.468){5}{\rule{1.600pt}{0.113pt}}
\multiput(433.00,828.17)(11.679,-4.000){2}{\rule{0.800pt}{0.400pt}}
\multiput(448.00,823.94)(2.236,-0.468){5}{\rule{1.700pt}{0.113pt}}
\multiput(448.00,824.17)(12.472,-4.000){2}{\rule{0.850pt}{0.400pt}}
\multiput(464.00,819.93)(1.601,-0.477){7}{\rule{1.300pt}{0.115pt}}
\multiput(464.00,820.17)(12.302,-5.000){2}{\rule{0.650pt}{0.400pt}}
\multiput(479.00,814.93)(1.601,-0.477){7}{\rule{1.300pt}{0.115pt}}
\multiput(479.00,815.17)(12.302,-5.000){2}{\rule{0.650pt}{0.400pt}}
\multiput(494.00,809.93)(1.304,-0.482){9}{\rule{1.100pt}{0.116pt}}
\multiput(494.00,810.17)(12.717,-6.000){2}{\rule{0.550pt}{0.400pt}}
\multiput(509.00,803.93)(1.304,-0.482){9}{\rule{1.100pt}{0.116pt}}
\multiput(509.00,804.17)(12.717,-6.000){2}{\rule{0.550pt}{0.400pt}}
\multiput(524.00,797.93)(0.956,-0.488){13}{\rule{0.850pt}{0.117pt}}
\multiput(524.00,798.17)(13.236,-8.000){2}{\rule{0.425pt}{0.400pt}}
\multiput(539.00,789.93)(0.902,-0.489){15}{\rule{0.811pt}{0.118pt}}
\multiput(539.00,790.17)(14.316,-9.000){2}{\rule{0.406pt}{0.400pt}}
\multiput(555.00,780.93)(0.844,-0.489){15}{\rule{0.767pt}{0.118pt}}
\multiput(555.00,781.17)(13.409,-9.000){2}{\rule{0.383pt}{0.400pt}}
\multiput(570.00,771.92)(0.625,-0.492){21}{\rule{0.600pt}{0.119pt}}
\multiput(570.00,772.17)(13.755,-12.000){2}{\rule{0.300pt}{0.400pt}}
\multiput(585.00,759.92)(0.576,-0.493){23}{\rule{0.562pt}{0.119pt}}
\multiput(585.00,760.17)(13.834,-13.000){2}{\rule{0.281pt}{0.400pt}}
\multiput(600.58,745.81)(0.494,-0.531){27}{\rule{0.119pt}{0.527pt}}
\multiput(599.17,746.91)(15.000,-14.907){2}{\rule{0.400pt}{0.263pt}}
\multiput(615.58,729.59)(0.494,-0.600){27}{\rule{0.119pt}{0.580pt}}
\multiput(614.17,730.80)(15.000,-16.796){2}{\rule{0.400pt}{0.290pt}}
\multiput(630.58,711.41)(0.494,-0.657){29}{\rule{0.119pt}{0.625pt}}
\multiput(629.17,712.70)(16.000,-19.703){2}{\rule{0.400pt}{0.313pt}}
\multiput(646.58,689.71)(0.494,-0.873){27}{\rule{0.119pt}{0.793pt}}
\multiput(645.17,691.35)(15.000,-24.353){2}{\rule{0.400pt}{0.397pt}}
\multiput(661.58,663.26)(0.494,-1.010){27}{\rule{0.119pt}{0.900pt}}
\multiput(660.17,665.13)(15.000,-28.132){2}{\rule{0.400pt}{0.450pt}}
\multiput(676.58,632.71)(0.494,-1.181){27}{\rule{0.119pt}{1.033pt}}
\multiput(675.17,634.86)(15.000,-32.855){2}{\rule{0.400pt}{0.517pt}}
\multiput(691.58,596.82)(0.494,-1.455){27}{\rule{0.119pt}{1.247pt}}
\multiput(690.17,599.41)(15.000,-40.412){2}{\rule{0.400pt}{0.623pt}}
\multiput(706.58,553.50)(0.494,-1.554){29}{\rule{0.119pt}{1.325pt}}
\multiput(705.17,556.25)(16.000,-46.250){2}{\rule{0.400pt}{0.663pt}}
\multiput(722.58,503.28)(0.494,-1.934){27}{\rule{0.119pt}{1.620pt}}
\multiput(721.17,506.64)(15.000,-53.638){2}{\rule{0.400pt}{0.810pt}}
\multiput(737.58,445.61)(0.494,-2.139){27}{\rule{0.119pt}{1.780pt}}
\multiput(736.17,449.31)(15.000,-59.306){2}{\rule{0.400pt}{0.890pt}}
\multiput(752.58,382.28)(0.494,-2.241){27}{\rule{0.119pt}{1.860pt}}
\multiput(751.17,386.14)(15.000,-62.139){2}{\rule{0.400pt}{0.930pt}}
\multiput(767.58,316.61)(0.494,-2.139){27}{\rule{0.119pt}{1.780pt}}
\multiput(766.17,320.31)(15.000,-59.306){2}{\rule{0.400pt}{0.890pt}}
\multiput(782.58,254.83)(0.494,-1.763){27}{\rule{0.119pt}{1.487pt}}
\multiput(781.17,257.91)(15.000,-48.914){2}{\rule{0.400pt}{0.743pt}}
\multiput(797.58,205.06)(0.494,-1.073){29}{\rule{0.119pt}{0.950pt}}
\multiput(796.17,207.03)(16.000,-32.028){2}{\rule{0.400pt}{0.475pt}}
\multiput(813.00,173.92)(0.625,-0.492){21}{\rule{0.600pt}{0.119pt}}
\multiput(813.00,174.17)(13.755,-12.000){2}{\rule{0.300pt}{0.400pt}}
\multiput(828.00,163.58)(0.625,0.492){21}{\rule{0.600pt}{0.119pt}}
\multiput(828.00,162.17)(13.755,12.000){2}{\rule{0.300pt}{0.400pt}}
\multiput(843.58,175.00)(0.494,1.113){27}{\rule{0.119pt}{0.980pt}}
\multiput(842.17,175.00)(15.000,30.966){2}{\rule{0.400pt}{0.490pt}}
\multiput(858.58,208.00)(0.494,1.763){27}{\rule{0.119pt}{1.487pt}}
\multiput(857.17,208.00)(15.000,48.914){2}{\rule{0.400pt}{0.743pt}}
\multiput(873.58,260.00)(0.494,2.002){29}{\rule{0.119pt}{1.675pt}}
\multiput(872.17,260.00)(16.000,59.523){2}{\rule{0.400pt}{0.838pt}}
\multiput(889.58,323.00)(0.494,2.241){27}{\rule{0.119pt}{1.860pt}}
\multiput(888.17,323.00)(15.000,62.139){2}{\rule{0.400pt}{0.930pt}}
\multiput(904.58,389.00)(0.494,2.139){27}{\rule{0.119pt}{1.780pt}}
\multiput(903.17,389.00)(15.000,59.306){2}{\rule{0.400pt}{0.890pt}}
\multiput(919.58,452.00)(0.494,1.934){27}{\rule{0.119pt}{1.620pt}}
\multiput(918.17,452.00)(15.000,53.638){2}{\rule{0.400pt}{0.810pt}}
\multiput(934.58,509.00)(0.494,1.694){27}{\rule{0.119pt}{1.433pt}}
\multiput(933.17,509.00)(15.000,47.025){2}{\rule{0.400pt}{0.717pt}}
\multiput(949.58,559.00)(0.494,1.421){27}{\rule{0.119pt}{1.220pt}}
\multiput(948.17,559.00)(15.000,39.468){2}{\rule{0.400pt}{0.610pt}}
\multiput(964.58,601.00)(0.494,1.137){29}{\rule{0.119pt}{1.000pt}}
\multiput(963.17,601.00)(16.000,33.924){2}{\rule{0.400pt}{0.500pt}}
\multiput(980.58,637.00)(0.494,1.010){27}{\rule{0.119pt}{0.900pt}}
\multiput(979.17,637.00)(15.000,28.132){2}{\rule{0.400pt}{0.450pt}}
\multiput(995.58,667.00)(0.494,0.839){27}{\rule{0.119pt}{0.767pt}}
\multiput(994.17,667.00)(15.000,23.409){2}{\rule{0.400pt}{0.383pt}}
\multiput(1010.58,692.00)(0.494,0.737){27}{\rule{0.119pt}{0.687pt}}
\multiput(1009.17,692.00)(15.000,20.575){2}{\rule{0.400pt}{0.343pt}}
\multiput(1025.58,714.00)(0.494,0.600){27}{\rule{0.119pt}{0.580pt}}
\multiput(1024.17,714.00)(15.000,16.796){2}{\rule{0.400pt}{0.290pt}}
\multiput(1040.00,732.58)(0.497,0.494){29}{\rule{0.500pt}{0.119pt}}
\multiput(1040.00,731.17)(14.962,16.000){2}{\rule{0.250pt}{0.400pt}}
\multiput(1056.00,748.58)(0.576,0.493){23}{\rule{0.562pt}{0.119pt}}
\multiput(1056.00,747.17)(13.834,13.000){2}{\rule{0.281pt}{0.400pt}}
\multiput(1071.00,761.58)(0.684,0.492){19}{\rule{0.645pt}{0.118pt}}
\multiput(1071.00,760.17)(13.660,11.000){2}{\rule{0.323pt}{0.400pt}}
\multiput(1086.00,772.58)(0.756,0.491){17}{\rule{0.700pt}{0.118pt}}
\multiput(1086.00,771.17)(13.547,10.000){2}{\rule{0.350pt}{0.400pt}}
\multiput(1101.00,782.59)(0.844,0.489){15}{\rule{0.767pt}{0.118pt}}
\multiput(1101.00,781.17)(13.409,9.000){2}{\rule{0.383pt}{0.400pt}}
\multiput(1116.00,791.59)(1.103,0.485){11}{\rule{0.957pt}{0.117pt}}
\multiput(1116.00,790.17)(13.013,7.000){2}{\rule{0.479pt}{0.400pt}}
\multiput(1131.00,798.59)(1.179,0.485){11}{\rule{1.014pt}{0.117pt}}
\multiput(1131.00,797.17)(13.895,7.000){2}{\rule{0.507pt}{0.400pt}}
\multiput(1147.00,805.59)(1.304,0.482){9}{\rule{1.100pt}{0.116pt}}
\multiput(1147.00,804.17)(12.717,6.000){2}{\rule{0.550pt}{0.400pt}}
\multiput(1162.00,811.59)(1.601,0.477){7}{\rule{1.300pt}{0.115pt}}
\multiput(1162.00,810.17)(12.302,5.000){2}{\rule{0.650pt}{0.400pt}}
\multiput(1177.00,816.59)(1.601,0.477){7}{\rule{1.300pt}{0.115pt}}
\multiput(1177.00,815.17)(12.302,5.000){2}{\rule{0.650pt}{0.400pt}}
\multiput(1192.00,821.60)(2.090,0.468){5}{\rule{1.600pt}{0.113pt}}
\multiput(1192.00,820.17)(11.679,4.000){2}{\rule{0.800pt}{0.400pt}}
\multiput(1207.00,825.60)(2.236,0.468){5}{\rule{1.700pt}{0.113pt}}
\multiput(1207.00,824.17)(12.472,4.000){2}{\rule{0.850pt}{0.400pt}}
\multiput(1223.00,829.61)(3.141,0.447){3}{\rule{2.100pt}{0.108pt}}
\multiput(1223.00,828.17)(10.641,3.000){2}{\rule{1.050pt}{0.400pt}}
\multiput(1238.00,832.61)(3.141,0.447){3}{\rule{2.100pt}{0.108pt}}
\multiput(1238.00,831.17)(10.641,3.000){2}{\rule{1.050pt}{0.400pt}}
\multiput(1253.00,835.61)(3.141,0.447){3}{\rule{2.100pt}{0.108pt}}
\multiput(1253.00,834.17)(10.641,3.000){2}{\rule{1.050pt}{0.400pt}}
\put(1268,838.17){\rule{3.100pt}{0.400pt}}
\multiput(1268.00,837.17)(8.566,2.000){2}{\rule{1.550pt}{0.400pt}}
\put(1283,840.17){\rule{3.100pt}{0.400pt}}
\multiput(1283.00,839.17)(8.566,2.000){2}{\rule{1.550pt}{0.400pt}}
\put(1298,842.17){\rule{3.300pt}{0.400pt}}
\multiput(1298.00,841.17)(9.151,2.000){2}{\rule{1.650pt}{0.400pt}}
\put(1314,844.17){\rule{3.100pt}{0.400pt}}
\multiput(1314.00,843.17)(8.566,2.000){2}{\rule{1.550pt}{0.400pt}}
\put(1329,846.17){\rule{3.100pt}{0.400pt}}
\multiput(1329.00,845.17)(8.566,2.000){2}{\rule{1.550pt}{0.400pt}}
\put(1344,847.67){\rule{3.614pt}{0.400pt}}
\multiput(1344.00,847.17)(7.500,1.000){2}{\rule{1.807pt}{0.400pt}}
\put(1359,849.17){\rule{3.100pt}{0.400pt}}
\multiput(1359.00,848.17)(8.566,2.000){2}{\rule{1.550pt}{0.400pt}}
\put(1374,850.67){\rule{3.854pt}{0.400pt}}
\multiput(1374.00,850.17)(8.000,1.000){2}{\rule{1.927pt}{0.400pt}}
\put(1390,852.17){\rule{3.100pt}{0.400pt}}
\multiput(1390.00,851.17)(8.566,2.000){2}{\rule{1.550pt}{0.400pt}}
\put(1405,853.67){\rule{3.614pt}{0.400pt}}
\multiput(1405.00,853.17)(7.500,1.000){2}{\rule{1.807pt}{0.400pt}}
\put(1420,854.67){\rule{3.614pt}{0.400pt}}
\multiput(1420.00,854.17)(7.500,1.000){2}{\rule{1.807pt}{0.400pt}}
\sbox{\plotpoint}{\rule[-0.600pt]{1.200pt}{1.200pt}}%
\put(221,746){\usebox{\plotpoint}}
\put(221,742.51){\rule{3.614pt}{1.200pt}}
\multiput(221.00,743.51)(7.500,-2.000){2}{\rule{1.807pt}{1.200pt}}
\put(236,740.51){\rule{3.614pt}{1.200pt}}
\multiput(236.00,741.51)(7.500,-2.000){2}{\rule{1.807pt}{1.200pt}}
\put(251,738.01){\rule{3.614pt}{1.200pt}}
\multiput(251.00,739.51)(7.500,-3.000){2}{\rule{1.807pt}{1.200pt}}
\put(266,735.51){\rule{3.614pt}{1.200pt}}
\multiput(266.00,736.51)(7.500,-2.000){2}{\rule{1.807pt}{1.200pt}}
\put(281,733.01){\rule{3.854pt}{1.200pt}}
\multiput(281.00,734.51)(8.000,-3.000){2}{\rule{1.927pt}{1.200pt}}
\put(297,729.51){\rule{3.614pt}{1.200pt}}
\multiput(297.00,731.51)(7.500,-4.000){2}{\rule{1.807pt}{1.200pt}}
\put(312,726.01){\rule{3.614pt}{1.200pt}}
\multiput(312.00,727.51)(7.500,-3.000){2}{\rule{1.807pt}{1.200pt}}
\put(327,722.01){\rule{3.614pt}{1.200pt}}
\multiput(327.00,724.51)(7.500,-5.000){2}{\rule{1.807pt}{1.200pt}}
\put(342,717.51){\rule{3.614pt}{1.200pt}}
\multiput(342.00,719.51)(7.500,-4.000){2}{\rule{1.807pt}{1.200pt}}
\multiput(357.00,715.25)(1.301,-0.509){2}{\rule{3.300pt}{0.123pt}}
\multiput(357.00,715.51)(8.151,-6.000){2}{\rule{1.650pt}{1.200pt}}
\multiput(372.00,709.25)(1.471,-0.509){2}{\rule{3.500pt}{0.123pt}}
\multiput(372.00,709.51)(8.736,-6.000){2}{\rule{1.750pt}{1.200pt}}
\multiput(388.00,703.26)(0.883,-0.503){6}{\rule{2.550pt}{0.121pt}}
\multiput(388.00,703.51)(9.707,-8.000){2}{\rule{1.275pt}{1.200pt}}
\multiput(403.00,695.26)(0.779,-0.502){8}{\rule{2.300pt}{0.121pt}}
\multiput(403.00,695.51)(10.226,-9.000){2}{\rule{1.150pt}{1.200pt}}
\multiput(418.00,686.26)(0.698,-0.502){10}{\rule{2.100pt}{0.121pt}}
\multiput(418.00,686.51)(10.641,-10.000){2}{\rule{1.050pt}{1.200pt}}
\multiput(433.00,676.26)(0.534,-0.501){16}{\rule{1.685pt}{0.121pt}}
\multiput(433.00,676.51)(11.503,-13.000){2}{\rule{0.842pt}{1.200pt}}
\multiput(448.00,663.26)(0.465,-0.501){22}{\rule{1.500pt}{0.121pt}}
\multiput(448.00,663.51)(12.887,-16.000){2}{\rule{0.750pt}{1.200pt}}
\multiput(466.24,642.78)(0.501,-0.567){20}{\rule{0.121pt}{1.740pt}}
\multiput(461.51,646.39)(15.000,-14.389){2}{\rule{1.200pt}{0.870pt}}
\multiput(481.24,623.12)(0.501,-0.743){20}{\rule{0.121pt}{2.140pt}}
\multiput(476.51,627.56)(15.000,-18.558){2}{\rule{1.200pt}{1.070pt}}
\multiput(496.24,598.79)(0.501,-0.884){20}{\rule{0.121pt}{2.460pt}}
\multiput(491.51,603.89)(15.000,-21.894){2}{\rule{1.200pt}{1.230pt}}
\multiput(511.24,570.79)(0.501,-0.989){20}{\rule{0.121pt}{2.700pt}}
\multiput(506.51,576.40)(15.000,-24.396){2}{\rule{1.200pt}{1.350pt}}
\multiput(526.24,540.46)(0.501,-1.024){20}{\rule{0.121pt}{2.780pt}}
\multiput(521.51,546.23)(15.000,-25.230){2}{\rule{1.200pt}{1.390pt}}
\multiput(541.24,510.41)(0.501,-0.925){22}{\rule{0.121pt}{2.550pt}}
\multiput(536.51,515.71)(16.000,-24.707){2}{\rule{1.200pt}{1.275pt}}
\multiput(557.24,480.46)(0.501,-0.919){20}{\rule{0.121pt}{2.540pt}}
\multiput(552.51,485.73)(15.000,-22.728){2}{\rule{1.200pt}{1.270pt}}
\multiput(572.24,453.78)(0.501,-0.778){20}{\rule{0.121pt}{2.220pt}}
\multiput(567.51,458.39)(15.000,-19.392){2}{\rule{1.200pt}{1.110pt}}
\multiput(587.24,430.78)(0.501,-0.673){20}{\rule{0.121pt}{1.980pt}}
\multiput(582.51,434.89)(15.000,-16.890){2}{\rule{1.200pt}{0.990pt}}
\multiput(602.24,410.78)(0.501,-0.567){20}{\rule{0.121pt}{1.740pt}}
\multiput(597.51,414.39)(15.000,-14.389){2}{\rule{1.200pt}{0.870pt}}
\multiput(617.24,393.44)(0.501,-0.497){20}{\rule{0.121pt}{1.580pt}}
\multiput(612.51,396.72)(15.000,-12.721){2}{\rule{1.200pt}{0.790pt}}
\multiput(630.00,381.26)(0.575,-0.501){16}{\rule{1.777pt}{0.121pt}}
\multiput(630.00,381.51)(12.312,-13.000){2}{\rule{0.888pt}{1.200pt}}
\multiput(646.00,368.26)(0.633,-0.502){12}{\rule{1.936pt}{0.121pt}}
\multiput(646.00,368.51)(10.981,-11.000){2}{\rule{0.968pt}{1.200pt}}
\multiput(661.00,357.26)(0.698,-0.502){10}{\rule{2.100pt}{0.121pt}}
\multiput(661.00,357.51)(10.641,-10.000){2}{\rule{1.050pt}{1.200pt}}
\multiput(676.00,347.26)(0.779,-0.502){8}{\rule{2.300pt}{0.121pt}}
\multiput(676.00,347.51)(10.226,-9.000){2}{\rule{1.150pt}{1.200pt}}
\multiput(691.00,338.26)(1.027,-0.505){4}{\rule{2.871pt}{0.122pt}}
\multiput(691.00,338.51)(9.040,-7.000){2}{\rule{1.436pt}{1.200pt}}
\multiput(706.00,331.25)(1.471,-0.509){2}{\rule{3.500pt}{0.123pt}}
\multiput(706.00,331.51)(8.736,-6.000){2}{\rule{1.750pt}{1.200pt}}
\put(722,323.01){\rule{3.614pt}{1.200pt}}
\multiput(722.00,325.51)(7.500,-5.000){2}{\rule{1.807pt}{1.200pt}}
\put(737,318.51){\rule{3.614pt}{1.200pt}}
\multiput(737.00,320.51)(7.500,-4.000){2}{\rule{1.807pt}{1.200pt}}
\put(752,314.51){\rule{3.614pt}{1.200pt}}
\multiput(752.00,316.51)(7.500,-4.000){2}{\rule{1.807pt}{1.200pt}}
\put(767,311.51){\rule{3.614pt}{1.200pt}}
\multiput(767.00,312.51)(7.500,-2.000){2}{\rule{1.807pt}{1.200pt}}
\put(782,309.51){\rule{3.614pt}{1.200pt}}
\multiput(782.00,310.51)(7.500,-2.000){2}{\rule{1.807pt}{1.200pt}}
\put(797,308.01){\rule{3.854pt}{1.200pt}}
\multiput(797.00,308.51)(8.000,-1.000){2}{\rule{1.927pt}{1.200pt}}
\put(813,307.01){\rule{3.614pt}{1.200pt}}
\multiput(813.00,307.51)(7.500,-1.000){2}{\rule{1.807pt}{1.200pt}}
\put(828,307.01){\rule{3.614pt}{1.200pt}}
\multiput(828.00,306.51)(7.500,1.000){2}{\rule{1.807pt}{1.200pt}}
\put(843,308.01){\rule{3.614pt}{1.200pt}}
\multiput(843.00,307.51)(7.500,1.000){2}{\rule{1.807pt}{1.200pt}}
\put(858,309.51){\rule{3.614pt}{1.200pt}}
\multiput(858.00,308.51)(7.500,2.000){2}{\rule{1.807pt}{1.200pt}}
\put(873,311.51){\rule{3.854pt}{1.200pt}}
\multiput(873.00,310.51)(8.000,2.000){2}{\rule{1.927pt}{1.200pt}}
\put(889,314.01){\rule{3.614pt}{1.200pt}}
\multiput(889.00,312.51)(7.500,3.000){2}{\rule{1.807pt}{1.200pt}}
\put(904,318.01){\rule{3.614pt}{1.200pt}}
\multiput(904.00,315.51)(7.500,5.000){2}{\rule{1.807pt}{1.200pt}}
\put(919,323.01){\rule{3.614pt}{1.200pt}}
\multiput(919.00,320.51)(7.500,5.000){2}{\rule{1.807pt}{1.200pt}}
\multiput(934.00,330.24)(1.301,0.509){2}{\rule{3.300pt}{0.123pt}}
\multiput(934.00,325.51)(8.151,6.000){2}{\rule{1.650pt}{1.200pt}}
\multiput(949.00,336.24)(1.027,0.505){4}{\rule{2.871pt}{0.122pt}}
\multiput(949.00,331.51)(9.040,7.000){2}{\rule{1.436pt}{1.200pt}}
\multiput(964.00,343.24)(0.843,0.502){8}{\rule{2.433pt}{0.121pt}}
\multiput(964.00,338.51)(10.949,9.000){2}{\rule{1.217pt}{1.200pt}}
\multiput(980.00,352.24)(0.779,0.502){8}{\rule{2.300pt}{0.121pt}}
\multiput(980.00,347.51)(10.226,9.000){2}{\rule{1.150pt}{1.200pt}}
\multiput(995.00,361.24)(0.579,0.501){14}{\rule{1.800pt}{0.121pt}}
\multiput(995.00,356.51)(11.264,12.000){2}{\rule{0.900pt}{1.200pt}}
\multiput(1010.00,373.24)(0.534,0.501){16}{\rule{1.685pt}{0.121pt}}
\multiput(1010.00,368.51)(11.503,13.000){2}{\rule{0.842pt}{1.200pt}}
\multiput(1027.24,384.00)(0.501,0.497){20}{\rule{0.121pt}{1.580pt}}
\multiput(1022.51,384.00)(15.000,12.721){2}{\rule{1.200pt}{0.790pt}}
\multiput(1042.24,400.00)(0.501,0.531){22}{\rule{0.121pt}{1.650pt}}
\multiput(1037.51,400.00)(16.000,14.575){2}{\rule{1.200pt}{0.825pt}}
\multiput(1058.24,418.00)(0.501,0.673){20}{\rule{0.121pt}{1.980pt}}
\multiput(1053.51,418.00)(15.000,16.890){2}{\rule{1.200pt}{0.990pt}}
\multiput(1073.24,439.00)(0.501,0.778){20}{\rule{0.121pt}{2.220pt}}
\multiput(1068.51,439.00)(15.000,19.392){2}{\rule{1.200pt}{1.110pt}}
\multiput(1088.24,463.00)(0.501,0.884){20}{\rule{0.121pt}{2.460pt}}
\multiput(1083.51,463.00)(15.000,21.894){2}{\rule{1.200pt}{1.230pt}}
\multiput(1103.24,490.00)(0.501,0.989){20}{\rule{0.121pt}{2.700pt}}
\multiput(1098.51,490.00)(15.000,24.396){2}{\rule{1.200pt}{1.350pt}}
\multiput(1118.24,520.00)(0.501,1.059){20}{\rule{0.121pt}{2.860pt}}
\multiput(1113.51,520.00)(15.000,26.064){2}{\rule{1.200pt}{1.430pt}}
\multiput(1133.24,552.00)(0.501,0.925){22}{\rule{0.121pt}{2.550pt}}
\multiput(1128.51,552.00)(16.000,24.707){2}{\rule{1.200pt}{1.275pt}}
\multiput(1149.24,582.00)(0.501,0.848){20}{\rule{0.121pt}{2.380pt}}
\multiput(1144.51,582.00)(15.000,21.060){2}{\rule{1.200pt}{1.190pt}}
\multiput(1164.24,608.00)(0.501,0.743){20}{\rule{0.121pt}{2.140pt}}
\multiput(1159.51,608.00)(15.000,18.558){2}{\rule{1.200pt}{1.070pt}}
\multiput(1179.24,631.00)(0.501,0.602){20}{\rule{0.121pt}{1.820pt}}
\multiput(1174.51,631.00)(15.000,15.222){2}{\rule{1.200pt}{0.910pt}}
\multiput(1194.24,650.00)(0.501,0.497){20}{\rule{0.121pt}{1.580pt}}
\multiput(1189.51,650.00)(15.000,12.721){2}{\rule{1.200pt}{0.790pt}}
\multiput(1207.00,668.24)(0.624,0.501){14}{\rule{1.900pt}{0.121pt}}
\multiput(1207.00,663.51)(12.056,12.000){2}{\rule{0.950pt}{1.200pt}}
\multiput(1223.00,680.24)(0.633,0.502){12}{\rule{1.936pt}{0.121pt}}
\multiput(1223.00,675.51)(10.981,11.000){2}{\rule{0.968pt}{1.200pt}}
\multiput(1238.00,691.24)(0.779,0.502){8}{\rule{2.300pt}{0.121pt}}
\multiput(1238.00,686.51)(10.226,9.000){2}{\rule{1.150pt}{1.200pt}}
\multiput(1253.00,700.24)(1.027,0.505){4}{\rule{2.871pt}{0.122pt}}
\multiput(1253.00,695.51)(9.040,7.000){2}{\rule{1.436pt}{1.200pt}}
\multiput(1268.00,707.24)(1.027,0.505){4}{\rule{2.871pt}{0.122pt}}
\multiput(1268.00,702.51)(9.040,7.000){2}{\rule{1.436pt}{1.200pt}}
\put(1283,712.01){\rule{3.614pt}{1.200pt}}
\multiput(1283.00,709.51)(7.500,5.000){2}{\rule{1.807pt}{1.200pt}}
\put(1298,717.01){\rule{3.854pt}{1.200pt}}
\multiput(1298.00,714.51)(8.000,5.000){2}{\rule{1.927pt}{1.200pt}}
\put(1314,722.01){\rule{3.614pt}{1.200pt}}
\multiput(1314.00,719.51)(7.500,5.000){2}{\rule{1.807pt}{1.200pt}}
\put(1329,726.01){\rule{3.614pt}{1.200pt}}
\multiput(1329.00,724.51)(7.500,3.000){2}{\rule{1.807pt}{1.200pt}}
\put(1344,729.51){\rule{3.614pt}{1.200pt}}
\multiput(1344.00,727.51)(7.500,4.000){2}{\rule{1.807pt}{1.200pt}}
\put(1359,733.01){\rule{3.614pt}{1.200pt}}
\multiput(1359.00,731.51)(7.500,3.000){2}{\rule{1.807pt}{1.200pt}}
\put(1374,735.51){\rule{3.854pt}{1.200pt}}
\multiput(1374.00,734.51)(8.000,2.000){2}{\rule{1.927pt}{1.200pt}}
\put(1390,738.01){\rule{3.614pt}{1.200pt}}
\multiput(1390.00,736.51)(7.500,3.000){2}{\rule{1.807pt}{1.200pt}}
\put(1405,740.51){\rule{3.614pt}{1.200pt}}
\multiput(1405.00,739.51)(7.500,2.000){2}{\rule{1.807pt}{1.200pt}}
\put(1420,742.51){\rule{3.614pt}{1.200pt}}
\multiput(1420.00,741.51)(7.500,2.000){2}{\rule{1.807pt}{1.200pt}}
\sbox{\plotpoint}{\rule[-0.400pt]{0.800pt}{0.800pt}}%
\put(221,374){\usebox{\plotpoint}}
\put(221,370.84){\rule{3.614pt}{0.800pt}}
\multiput(221.00,372.34)(7.500,-3.000){2}{\rule{1.807pt}{0.800pt}}
\put(236,367.84){\rule{3.614pt}{0.800pt}}
\multiput(236.00,369.34)(7.500,-3.000){2}{\rule{1.807pt}{0.800pt}}
\put(251,364.84){\rule{3.614pt}{0.800pt}}
\multiput(251.00,366.34)(7.500,-3.000){2}{\rule{1.807pt}{0.800pt}}
\put(266,362.34){\rule{3.614pt}{0.800pt}}
\multiput(266.00,363.34)(7.500,-2.000){2}{\rule{1.807pt}{0.800pt}}
\put(281,359.84){\rule{3.854pt}{0.800pt}}
\multiput(281.00,361.34)(8.000,-3.000){2}{\rule{1.927pt}{0.800pt}}
\put(297,357.34){\rule{3.614pt}{0.800pt}}
\multiput(297.00,358.34)(7.500,-2.000){2}{\rule{1.807pt}{0.800pt}}
\put(312,355.34){\rule{3.614pt}{0.800pt}}
\multiput(312.00,356.34)(7.500,-2.000){2}{\rule{1.807pt}{0.800pt}}
\put(327,352.84){\rule{3.614pt}{0.800pt}}
\multiput(327.00,354.34)(7.500,-3.000){2}{\rule{1.807pt}{0.800pt}}
\put(342,350.34){\rule{3.614pt}{0.800pt}}
\multiput(342.00,351.34)(7.500,-2.000){2}{\rule{1.807pt}{0.800pt}}
\put(357,348.84){\rule{3.614pt}{0.800pt}}
\multiput(357.00,349.34)(7.500,-1.000){2}{\rule{1.807pt}{0.800pt}}
\put(372,347.34){\rule{3.854pt}{0.800pt}}
\multiput(372.00,348.34)(8.000,-2.000){2}{\rule{1.927pt}{0.800pt}}
\put(388,345.34){\rule{3.614pt}{0.800pt}}
\multiput(388.00,346.34)(7.500,-2.000){2}{\rule{1.807pt}{0.800pt}}
\put(403,343.34){\rule{3.614pt}{0.800pt}}
\multiput(403.00,344.34)(7.500,-2.000){2}{\rule{1.807pt}{0.800pt}}
\put(418,341.84){\rule{3.614pt}{0.800pt}}
\multiput(418.00,342.34)(7.500,-1.000){2}{\rule{1.807pt}{0.800pt}}
\put(433,340.34){\rule{3.614pt}{0.800pt}}
\multiput(433.00,341.34)(7.500,-2.000){2}{\rule{1.807pt}{0.800pt}}
\put(448,338.84){\rule{3.854pt}{0.800pt}}
\multiput(448.00,339.34)(8.000,-1.000){2}{\rule{1.927pt}{0.800pt}}
\put(464,337.84){\rule{3.614pt}{0.800pt}}
\multiput(464.00,338.34)(7.500,-1.000){2}{\rule{1.807pt}{0.800pt}}
\put(479,336.34){\rule{3.614pt}{0.800pt}}
\multiput(479.00,337.34)(7.500,-2.000){2}{\rule{1.807pt}{0.800pt}}
\put(494,334.84){\rule{3.614pt}{0.800pt}}
\multiput(494.00,335.34)(7.500,-1.000){2}{\rule{1.807pt}{0.800pt}}
\put(509,333.84){\rule{3.614pt}{0.800pt}}
\multiput(509.00,334.34)(7.500,-1.000){2}{\rule{1.807pt}{0.800pt}}
\put(524,332.84){\rule{3.614pt}{0.800pt}}
\multiput(524.00,333.34)(7.500,-1.000){2}{\rule{1.807pt}{0.800pt}}
\put(539,331.84){\rule{3.854pt}{0.800pt}}
\multiput(539.00,332.34)(8.000,-1.000){2}{\rule{1.927pt}{0.800pt}}
\put(555,330.84){\rule{3.614pt}{0.800pt}}
\multiput(555.00,331.34)(7.500,-1.000){2}{\rule{1.807pt}{0.800pt}}
\put(570,329.84){\rule{3.614pt}{0.800pt}}
\multiput(570.00,330.34)(7.500,-1.000){2}{\rule{1.807pt}{0.800pt}}
\put(600,328.84){\rule{3.614pt}{0.800pt}}
\multiput(600.00,329.34)(7.500,-1.000){2}{\rule{1.807pt}{0.800pt}}
\put(615,327.84){\rule{3.614pt}{0.800pt}}
\multiput(615.00,328.34)(7.500,-1.000){2}{\rule{1.807pt}{0.800pt}}
\put(630,326.84){\rule{3.854pt}{0.800pt}}
\multiput(630.00,327.34)(8.000,-1.000){2}{\rule{1.927pt}{0.800pt}}
\put(585.0,331.0){\rule[-0.400pt]{3.613pt}{0.800pt}}
\put(661,325.84){\rule{3.614pt}{0.800pt}}
\multiput(661.00,326.34)(7.500,-1.000){2}{\rule{1.807pt}{0.800pt}}
\put(646.0,328.0){\rule[-0.400pt]{3.613pt}{0.800pt}}
\put(691,324.84){\rule{3.614pt}{0.800pt}}
\multiput(691.00,325.34)(7.500,-1.000){2}{\rule{1.807pt}{0.800pt}}
\put(676.0,327.0){\rule[-0.400pt]{3.613pt}{0.800pt}}
\put(752,323.84){\rule{3.614pt}{0.800pt}}
\multiput(752.00,324.34)(7.500,-1.000){2}{\rule{1.807pt}{0.800pt}}
\put(706.0,326.0){\rule[-0.400pt]{11.081pt}{0.800pt}}
\put(889,323.84){\rule{3.614pt}{0.800pt}}
\multiput(889.00,323.34)(7.500,1.000){2}{\rule{1.807pt}{0.800pt}}
\put(767.0,325.0){\rule[-0.400pt]{29.390pt}{0.800pt}}
\put(949,324.84){\rule{3.614pt}{0.800pt}}
\multiput(949.00,324.34)(7.500,1.000){2}{\rule{1.807pt}{0.800pt}}
\put(904.0,326.0){\rule[-0.400pt]{10.840pt}{0.800pt}}
\put(980,325.84){\rule{3.614pt}{0.800pt}}
\multiput(980.00,325.34)(7.500,1.000){2}{\rule{1.807pt}{0.800pt}}
\put(964.0,327.0){\rule[-0.400pt]{3.854pt}{0.800pt}}
\put(1010,326.84){\rule{3.614pt}{0.800pt}}
\multiput(1010.00,326.34)(7.500,1.000){2}{\rule{1.807pt}{0.800pt}}
\put(1025,327.84){\rule{3.614pt}{0.800pt}}
\multiput(1025.00,327.34)(7.500,1.000){2}{\rule{1.807pt}{0.800pt}}
\put(1040,328.84){\rule{3.854pt}{0.800pt}}
\multiput(1040.00,328.34)(8.000,1.000){2}{\rule{1.927pt}{0.800pt}}
\put(995.0,328.0){\rule[-0.400pt]{3.613pt}{0.800pt}}
\put(1071,329.84){\rule{3.614pt}{0.800pt}}
\multiput(1071.00,329.34)(7.500,1.000){2}{\rule{1.807pt}{0.800pt}}
\put(1086,330.84){\rule{3.614pt}{0.800pt}}
\multiput(1086.00,330.34)(7.500,1.000){2}{\rule{1.807pt}{0.800pt}}
\put(1101,331.84){\rule{3.614pt}{0.800pt}}
\multiput(1101.00,331.34)(7.500,1.000){2}{\rule{1.807pt}{0.800pt}}
\put(1116,332.84){\rule{3.614pt}{0.800pt}}
\multiput(1116.00,332.34)(7.500,1.000){2}{\rule{1.807pt}{0.800pt}}
\put(1131,333.84){\rule{3.854pt}{0.800pt}}
\multiput(1131.00,333.34)(8.000,1.000){2}{\rule{1.927pt}{0.800pt}}
\put(1147,334.84){\rule{3.614pt}{0.800pt}}
\multiput(1147.00,334.34)(7.500,1.000){2}{\rule{1.807pt}{0.800pt}}
\put(1162,336.34){\rule{3.614pt}{0.800pt}}
\multiput(1162.00,335.34)(7.500,2.000){2}{\rule{1.807pt}{0.800pt}}
\put(1177,337.84){\rule{3.614pt}{0.800pt}}
\multiput(1177.00,337.34)(7.500,1.000){2}{\rule{1.807pt}{0.800pt}}
\put(1192,338.84){\rule{3.614pt}{0.800pt}}
\multiput(1192.00,338.34)(7.500,1.000){2}{\rule{1.807pt}{0.800pt}}
\put(1207,340.34){\rule{3.854pt}{0.800pt}}
\multiput(1207.00,339.34)(8.000,2.000){2}{\rule{1.927pt}{0.800pt}}
\put(1223,341.84){\rule{3.614pt}{0.800pt}}
\multiput(1223.00,341.34)(7.500,1.000){2}{\rule{1.807pt}{0.800pt}}
\put(1238,343.34){\rule{3.614pt}{0.800pt}}
\multiput(1238.00,342.34)(7.500,2.000){2}{\rule{1.807pt}{0.800pt}}
\put(1253,345.34){\rule{3.614pt}{0.800pt}}
\multiput(1253.00,344.34)(7.500,2.000){2}{\rule{1.807pt}{0.800pt}}
\put(1268,347.34){\rule{3.614pt}{0.800pt}}
\multiput(1268.00,346.34)(7.500,2.000){2}{\rule{1.807pt}{0.800pt}}
\put(1283,348.84){\rule{3.614pt}{0.800pt}}
\multiput(1283.00,348.34)(7.500,1.000){2}{\rule{1.807pt}{0.800pt}}
\put(1298,350.34){\rule{3.854pt}{0.800pt}}
\multiput(1298.00,349.34)(8.000,2.000){2}{\rule{1.927pt}{0.800pt}}
\put(1314,352.84){\rule{3.614pt}{0.800pt}}
\multiput(1314.00,351.34)(7.500,3.000){2}{\rule{1.807pt}{0.800pt}}
\put(1329,355.34){\rule{3.614pt}{0.800pt}}
\multiput(1329.00,354.34)(7.500,2.000){2}{\rule{1.807pt}{0.800pt}}
\put(1344,357.34){\rule{3.614pt}{0.800pt}}
\multiput(1344.00,356.34)(7.500,2.000){2}{\rule{1.807pt}{0.800pt}}
\put(1359,359.34){\rule{3.614pt}{0.800pt}}
\multiput(1359.00,358.34)(7.500,2.000){2}{\rule{1.807pt}{0.800pt}}
\put(1374,361.84){\rule{3.854pt}{0.800pt}}
\multiput(1374.00,360.34)(8.000,3.000){2}{\rule{1.927pt}{0.800pt}}
\put(1390,364.84){\rule{3.614pt}{0.800pt}}
\multiput(1390.00,363.34)(7.500,3.000){2}{\rule{1.807pt}{0.800pt}}
\put(1405,367.84){\rule{3.614pt}{0.800pt}}
\multiput(1405.00,366.34)(7.500,3.000){2}{\rule{1.807pt}{0.800pt}}
\put(1420,370.84){\rule{3.614pt}{0.800pt}}
\multiput(1420.00,369.34)(7.500,3.000){2}{\rule{1.807pt}{0.800pt}}
\put(1056.0,331.0){\rule[-0.400pt]{3.613pt}{0.800pt}}
\end{picture}